\documentclass[fleqn,twoside]{article}
\usepackage{espcrc2}
\usepackage{graphicx}

\usepackage{amsmath, amssymb, amsfonts, mathrsfs}
\usepackage{epsfig}

\title{A Short Review of Time Dependent Solutions and Space-like Singularities in
String Theory}

\author{Micha Berkooz and Dori Reichmann\address[MCSD]{Department of Particle Physics,
The Weizmann Institute of Science, Rehovot 76100, Israel. \\}
\thanks{micha.berkooz@weizmann.ac.il, dor.reichmann@weizmann.ac.il}}

\begin{document}

\begin{abstract}
These lecture notes provide a short review of the status of time
dependent backgrounds in String theory, and in particular those that
contain space-like singularities. Despite considerable efforts, we
do not have yet a full and compelling picture of such backgrounds.
We review some of the various attempts to understand these
singularities via generalizations of the BKL dynamics, using
worldsheet methods and using non-perturbative tools such as the
AdS/CFT correspondence and M(atrix) theory. These lecture notes are
based on talks given at Cargese 06 and the dead-sea conference 06.
 \vspace{1pc}
\end{abstract}

\maketitle

\section{Introduction}

Over the last couple of decades string theory has provided us with a
continuous stream of new ideas and insight into general relativity.
Examples include the resolution of time-like singularities, either
by perturbative or non-perturbative techniques, our detailed
understanding of a large class of black holes, the AdS/CFT
correspondence and others. Motivated by this success and by
improved, observation driven, understanding of the evolution of our
universe, it is very tempting to try and study time dependent
solutions in String theory. This lecture will focus on a subset of
such recent attempts, and in particular on backgrounds with
space-like singularities and/or closed time like curves (CTC), which
exist in many time-dependent backgrounds.

Despite considerable effort over the last few years, it is hard to
say that we have a complete and compelling story about such
backgrounds. However, some hints are beginning to emerge about the
perturbative and non-perturbative effects that may play a role. This
talk will be a brief review of some of these hints.

Due to the lack of space and time, the list of topics that will be
covered will be partial, and many relevant and important
developments will not be touched upon. The latter include the
construction of big-bang singularities in AdS using "designer
gravity"
\cite{Hertog:2005hu,Hertog:2004ns,Hertog:2004jx,Hertog:2004rz}, the
"stiff equation of state" approach to big-bang singularities
\cite{Banks:2006hy,Banks:2004cw,Banks:2002fe,Banks:2001px}, String
gas cosmology (\cite{Biswas:2006bs} and references therein),
pre-big-bang cosmology (for a recent review see
\cite{Gasperini:2007vw}) and others. We will also not touch upon the
implications to cosmology (which is one of the driving forces behind
the study of time dependent backgrounds - see \cite{Khoury:2001wf}
and subsequent work). We apologize in advance for these
deficiencies.

The outline of this talk is the following. In section 2 we review
some of the stringy backgrounds that we will use later on. These
include primarily time-dependent orbifolds. In section 3 we review
the BKL dynamics, which is the generic approach to a spacelike
singularity in GR. In section 4 we discuss the perturbative
structure of Misner space, which is a prototype of stringy
space-like singularities. In section 5 we discuss another
perturbative technique, which is tachyon condensation. In section 6
we discuss non-perturbative approaches to spacelike singularities
via the AdS/CFT correspondence and M(atrix) theory.

\section{Examples of stringy time-dependent solutions}

\subsection{The Schwarzschild black hole}

\begin{figure}[ht]
\begin{center}
\includegraphics[width=7cm]{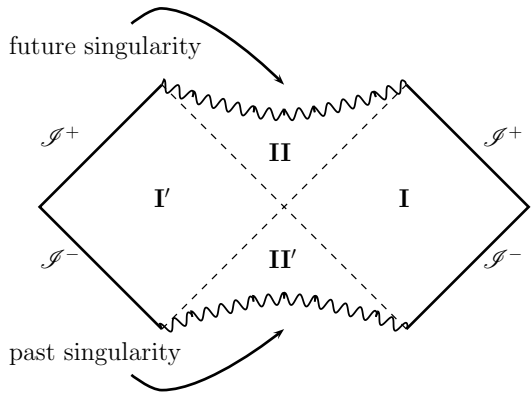}
\caption{\label{schwarz} Penrose diagram for the Kruskal extension
of the Schwarzschild black hole, showing conformal infinity as well
as the two singularities.}
\end{center}
\end{figure}

Probably the most common example of a space-like singularity is the
one in the interior of a Schwarzschild black hole. The Penrose
diagram of the eternal Schwarzschild black hole is given in figure
\ref{schwarz}. Regions I and I' are the two causally disconnected
'outside the horizon' regions. Each of them asymptotes to flat
Minkowski space far away from the black hole.
%at the conformal future and past
%boundaries ($\mathscr{I}^+$, $\mathscr{I}^-$).
Regions II,II' are the past and future 'inside the horizon' regions
(the horizons are the diagonal lines). Region II (II') ends (begins)
in a spacelike future (past) singularity. The Penrose diagram, which
appears in figure (1), is a convenient tool to encode the causal
structure of spacetime as all particles move within (or on the
boundaries) of the forward lightcone in the diagram.

The singularity inside the blackhole is closely related to the issue
of time-dependence, since the solution inside the black hole is
time-dependent. This is sometimes phrased as the question "what does
the in-falling observer see as he/she evolves towards the
singularity".

%%%%%%%%%%%%%%%%%%%%%%%%%%%%%%%%%%%%%%%%%%%%%%%%%%%%%%%%%%%%%%%%%%%%%%%%%%%%%%%%%%%%%5
%%%%%%%%%%%%%%%%%%%%%%%%%%%%%%%%%%%%%%%%%%%%%%%%%%%%%%%%%%%%%%%%%%%%%%%%%%%%%%%%%%%%%5
%%%%%%%%%%%%%%%%%%%%%%%%%%%%%%%%%%%%%%%%%%%%%%%%%%%%%%%%%%%%%%%%%%%%%%%%%%%%%%%%%%%%%5

\subsection{Black holes in $AdS$ and the BTZ black hole}

\begin{figure}[ht]
\begin{center}
\includegraphics[width=7cm]{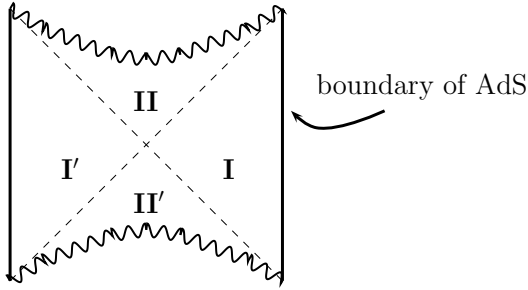}
\caption{\label{adsbh} The AdS$_d$, $d>3$, eternal black hole, the
causal structure is similar to the Schwarzschild black hole only
with the asymptotic boundary changed.}
\end{center}
\end{figure}

The AdS analog of the Schwarzschild solution is the AdS eternal
black hole. The casual structure of the solution
\cite{Fidkowski:2003nf}, given in the Penrose diagram in figure
\ref{adsbh}, is reminiscent of the Schwarzschild solution. The main
difference is the change of the asymptotic geometry from that of
flat space to that of AdS.

In $AdS_3$ the black hole is the BTZ black hole
\cite{Banados:1992gq,Banados:1992wn}. Its causal structure is
actually slightly different than figure \ref{adsbh} - for a
discussion of this issue in the context of space-like singularities
see \cite{Kraus:2002iv}. The metric of the non-rotating BTZ black
hole is (setting the AdS$_3$ radius to 1):
\begin{gather}
    ds^2 = -(r^2-M)dt^2+\frac{dr^2}{r^2-M}+r^2d\phi^2 \cr
    \phi\cong \phi+2\pi
\end{gather}

Alternatively, we can use the fact that AdS$_3$ is the group
manifold SL$(2,\mathbb{R})$ (more precisely, one may need to go to a
cover of SL(2) in the Minkowski case):
\begin{align}
    g\in\mathrm{SL}(2,\mathbb{R})&&
    ds^2 = \mathrm{Tr}\bigl(g^{-1}dgg^{-1}d g\bigr).
\end{align}
In String theory, $AdS_3$ with NS-NS fluxes can be promoted to a
full string theory solution (for example \cite{Giveon:1998ns} and
subsequent work) by taking an SL$(2,\mathbb{R})$-WZW model, which in
addition to a non-trivial metric also includes an $H=dB\not=0$ field
on $AdS_3$. The non-rotating BTZ black hole is then obtained by
orbifolding the geometry by an hyperbolic element of the isometry,
\begin{align}
    g\cong e^{\pi \sqrt{M}\sigma^3}\,g\,e^{\pi \sqrt{M}\sigma^3} && \sigma^3=\begin{pmatrix} 1 & 0 \\ 0 & -1\end{pmatrix}
\end{align}

Although the BTZ space-time contains a singularity and an event
horizon, the BTZ black hole geometry has constant curvature
(inherited from the AdS$_3$ geometry), which is very low for a large
$AdS$. The BTZ black hole might therefore be a suitable laboratory
to study the singularity inside a black hole, in which we can
disentangle the strong curvature effects from the effects of the
spacelike singularity or CTC's.

\subsection{Misner and Grant spaces}

\begin{figure}[ht]
\begin{center}
\includegraphics[width=7cm]{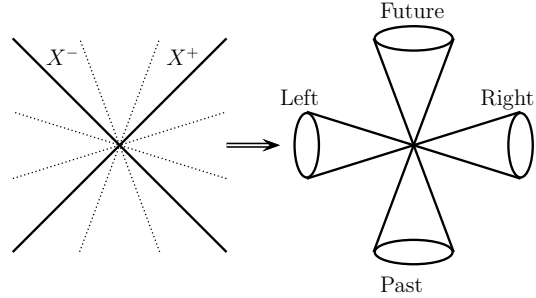}
\caption{\label{misner} The Minkowski space time (on the left) is
divided into 4 region by the action of the boost orbifold. The
dotted lines are the boundaries of the orbifold's fundamental
region, and are identified. The boost orbifold geometry (on the
right) is that of 4 cones. Each quadrant of Minkowski spacetime is
mapped in the orbifold to one of the orbifold's cones.}
\end{center}
\end{figure}

In analogy to the BTZ black hole, we wish to construct a laboratory
for big-bang physics. The simplest example is Misner space,
alternatively the orbifold ~$\mathbb{R}^{1,1}\bigm/boost$
\cite{Horowitz:1991ap,Nekrasov:2002kf,Pioline:2003bs}. To define the
orbifold one first writes Minkowski spacetime in lightcone
coordinates,
\begin{gather}
    ds^2 = -dt^2+dx^2 = -dX^+dX^-
    \cr
    X^\pm = t \pm x
\end{gather}
and then one mods out by the $Z$-orbifold
\begin{equation}
    (X^+,X^-)\rightarrow (e^{n\beta}X^+,e^{-n\beta}X^-),\ \ \ n\in
    \mathbb{Z}
\end{equation}
The action of the orbifold divides spacetime into 4 two-dimensional
regions as in figure \ref{misner} (there are 4 additional
one-dimensional regions on the lightcone but they will not play a
role below). Regions '\emph{future}' (F) and '\emph{past}' (P), are
the folding of the future and past cones (of Minkowski space) by a
spacelike identification. These regions are reminiscent of a
big-bang and a big-crunch regions of a cosmological model, which
becomes clear if we go to coordinates:
\begin{align}
\label{coora}
    &T\equiv \sqrt{X^+X^-}&
    &\theta \equiv \frac12\log\frac{X^+}{X^-}&
    \cr
%    ds^2=-2d\left(Te^{R}\right)d\left(Te^{-R}\right)
    &ds^2=-dT^2+T^2d\theta^2&
    &\theta\cong \theta+\beta&
\end{align}
Regions '\emph{left}' (L) and '\emph{right}' (R), are the folding of
the left and right cones (of Minkowski space) by a timelike
identification. These regions contain closed timelike curves (CTC)
since the generator of the boost is time-like there. A convenient
set of coordinates is
\begin{align}
\label{coorb}
    &R\equiv \sqrt{|X^+X^-|}&
    &\eta \equiv \frac12\log\left|\frac{X^+}{X^-}\right|&
    \cr
    &ds^2=-R^2d\eta^2+dR^2&
    &\eta\cong \eta+\beta&
\end{align}
We will refer to these regions as "whiskers".
A related example is the shifted boost orbifold (also known as Grant
space, and discussed extensively in
\cite{Cornalba:2003tq,Cornalba:2003ze,Cornalba:2002nv,Cornalba:2002fi}).
This space is defined by the addition, to the boost action, of a
shift in a transverse coordinate. In terms of Minkowskian
coordinates the action of the orbifold is
\begin{gather}
\begin{split}
    (X^+,X^-,X)\rightarrow
    e^{n{\hat\zeta}}(X^+,X^-,X)=\\
    =(e^{n\beta}X^+.e^{-n\beta}X^-,X+n\Delta)
\end{split}\cr
%    \begin{pmatrix}X^+\\ X^- \\ X\end{pmatrix}\cong e^{n\hat\zeta}\begin{pmatrix}X^+\\X^-\\X\end{pmatrix}
%   =\begin{pmatrix}e^{n\beta}X^+\\e^{-n\beta}X^-\\X+n\Delta\end{pmatrix}
%    \cr
%    n\in\mathbb{Z}\cr
    \hat\zeta = \beta\left(X^+\partial_{X^+}-X^-\partial_{X^-}\right)+\Delta\partial_{X}\cr
    n\in\mathbb{Z}
\end{gather}
Due to the shift of $X$, the orbifold does not posses any fixed
point and therefore leads to a smooth spacetime, although it still
contains CTCs.

\begin{figure}[ht]
\begin{center}
\includegraphics[width=7cm]{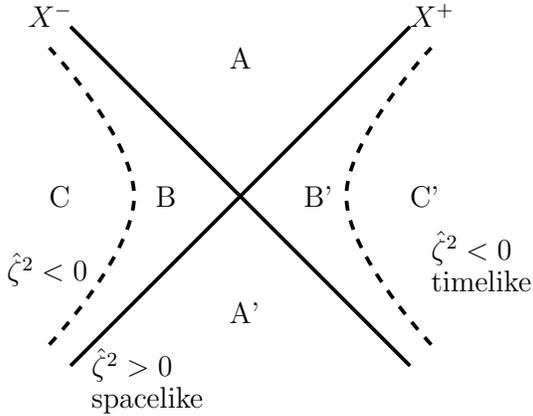}
\caption{\label{shift} Minkowski space is divided by the shifted
boost orbifold into 6 regions (seen at a fixed X cross section). The
dashed curves is where $\hat\zeta^2=0$.}
\end{center}
\end{figure}

Although Grant space is smooth, one can still distinguish different
regions with distinct features. There are 6 regions, which are
depicted in figure \ref{shift} by projecting them to the $X^+-X^-$
plane at $X=0$. The 'new' region (compared to the boost orbifold)
are regions B,B'. In the boost orbifold (without a shift) the
Killing vector ($\hat\zeta$) used for the identification was
spacelike in the future/past regions, timelike in the left/right
region and lightlike on the lightcone. In the shifted boost case
$\hat\zeta^2=0$ is a curve which lies in the left/right quadrants of
Minkowski space dividing them into regions B,B' where $\hat\zeta$ is
spacelike and regions C,C' with $\hat\zeta$ is timelike. Both
regions B,C (and B',C') contains CTC. However, all CTCs must pass
inside regions C or C', thus we can attempt to cut regions C,C' from
the space-time and remove all CTCs.

\subsection{The null brane}

Our last example is the null orbifold also known as the null-brane
\cite{Liu:2002kb,Liu:2002ft,Simon:2002ma}. The Lorentz group in 1+2
dimensions has three classes. A Killing vector from the elliptic
class (a spacelike rotation) generates the $\mathbb{Z}_n$ orbifolds,
a Killing vector from the hyperbolic class (i.e a boost) generates
the boost orbifold (Misner space) and a Killing vector in the
parabolic class generates the null orbifold. If one starts with a
supersymmetric theory on Minkowski space, the null orbifold is the
only one which will not break supersymmetry.

In terms of Minkowskian coordinates the null orbifold is defined as:
\begin{gather}
    \begin{split}
    \begin{pmatrix}X^+\\ X^- \\ X\end{pmatrix}\cong & e^{\nu\hat{\mathcal{J}}}\begin{pmatrix}X^+\\X^-\\X\end{pmatrix}=\cr
    =&\begin{pmatrix}X^+\\X^-+\nu X+\frac{1}{2}\nu^2X^+\\X+\nu X^+\end{pmatrix}
    \end{split}
    \cr
    \nu=2\pi n\quad n\in\mathbb{Z}\cr
    \hat{\mathcal{J}}= X^+\partial_{X}+X\partial_{X^-}.
\end{gather}
It is convenient to describe the geometry of the orbifold by
introducing new coordinates
\begin{align}
    Y^+=X^+&&
    Y=\frac{X}{X^+}&&
    Y^-=X^--\frac{X^2}{2X^+}
\end{align}
In terms of these coordinates the identification and the metric are
simple
\begin{gather}
    \left(Y^+,Y^-,Y\right)\cong\left(Y^+,Y^-,Y+\nu\right)
    \cr
    ds^2 = -2dY^+dY^-+(Y^+)^2(dY)^2
\end{gather}
This spacetime (also called the parabolic pinch) may be visualized
as two cones (parameterized by $Y^+$, and $Y$) with a common tip at
$Y^+=0$, times a real line (the $Y^-$ coordinate). $Y$ plays the
role of an "angular variables" of the cones, and $Y^+$ plays the
role of a "radial coordinate". As a function of the 'light-cone
time' $Y^+$ we have a big crunch of the circle at $Y^+=0$ which is
followed by a big bang. The dual role of $Y^+$ as both a radial
variable and a time variable is the source of some interesting
physics.

The down side of using the parabolic pinch coordinates is the
description of the orbifold at $X^+=0$. In terms of the Minkowskian
coordinates we identify that at the plane $X^+=0$ contains two
co-dimension 1 cones with a common tip at $X=0$.

\subsection{Relation between the models}

\begin{figure}[ht]
\begin{center}
\includegraphics[width=7cm]{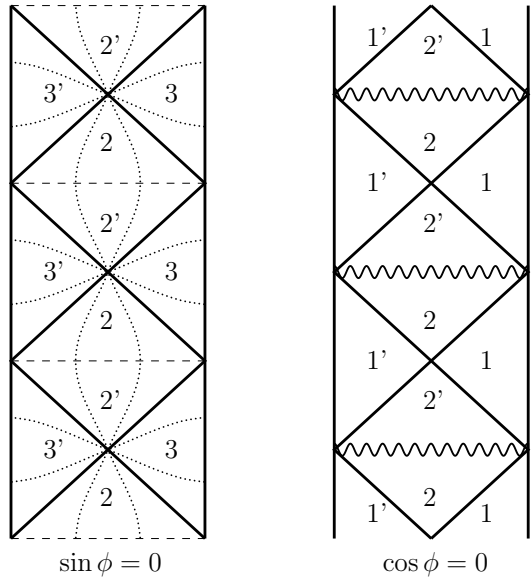}
\caption{\label{BTZ} The geometry of the BTZ black hole (with no
angular momentum) as seen from two different cross section. The
cross-section on the right is the same viewpoint described earlier
(figure \ref{adsbh}) where we see the outside the horizon area
(1,1') including the asymptotic boundary, the singularities and the
inner horizon area's (2,2'). In the cross-section on the left we see
the inner horizon areas (2,2') but also the behind the singularity
areas (3,3')}
\end{center}
\end{figure}

We have motivated the orbifolds of 3-dim Minkowski space as
laboratories for cosmological models, but we can also view them as
"local models" for the behavior inside the horizon of the BTZ black
holes. The structure of the non-rotating BTZ black hole is given in
figure \ref{BTZ}, which contains two slices of global Minkowski
$AdS_3$ (prior to the identification). Both slices contain the time
direction and an additional spatial directions. The spatial
directions of the two slices are at 90 degrees to each other. In
these diagrams regions 1 and 1' are normal regions outside the
horizon. Regions 2 and 2' are regions between the horizon and the
singularity, and regions 3 and 3' are behind the singularity.
Regions 2,2' are like P and F, and regions 3 and 3' are like L and
R. They will become precisely that in the limit that the radius of
curvature of $AdS_3$ is taken to infinity, while keep the distance
to the singularity fixed. The mass of the black hole is translated
to $\beta$. The addition of angular momentum to the black hole
deforms the geometry such that in the large radius limit, the "near
singularity" geometry is exactly that of the shifted boost orbifold,
the angular momentum is translated to the shift parameter
($\Delta$). Finally if we take a double scaling limit such that
$M=J\rightarrow 0$ as we take the AdS radius to infinity we find in
the near singularity are the geometry of the null brane.

\smallskip

There are many other models which describe string theory in time
dependent backgrounds, for example
\cite{Giveon:2003ge,Giveon:2003gb,Elitzur:2002vw,Elitzur:2002rt,Balasubramanian:2002ry,Biswas:2003ku}
and many others. Some of the features there are similar to the ones
that we will discuss here, but there are certainly many more
interesting aspects in the different models. We will not have time
to survey them.

\section{BKL dynamics}

The BKL (Belinsky-Khalatnikov-Lifshitz)
\cite{Belinsky:1970ew,Belinsky:1982pk} dynamics is considered to be
the generic behavior in general relativity (with potentially
additional fields) near a space-like singularity. We will begin by
reviewing the BKL analysis, which is a completely solvable model
based on a homogeneous 3-dimensional spatial slice which collapses
towards the singularity, and postpone the (less rigorous) discussion
of the non-homogeneous case to later.

The starting point is the Kasner family of solutions. In this class
of solutions one considers a 3 torus whose radii are allowed to vary
with time, i.e.
\begin{equation}
\label{kasnera}
 ds^2=-dt^2+\sum_{i=1}^3 a_i^2(t) dx_i^2
\end{equation}
where $x_i$ are periodic variables. A solution to Einstein's
equation can be obtained by setting
\begin{equation}
\label{kasnerb} a_i^2\sim t^{2p_i},\ \Sigma p_i=\Sigma p_i^2=1
\end{equation}
(multiplication of the $a_i$ by constants can be absorbed in the
radius of $x_i$). Misner space, which we have discussed before, is a
specific case of the Kasner solution in which $p_1=1$ and
$p_2=p_3=0$. More generally it can be shown that the solutions to
the constraints (\ref{kasnerb}) are such that if ordered
$p_1<p_2<p_3$ then
\begin{equation}
    -\frac13\leq p_1\leq 0\leq p_2\leq \frac23\leq p_3 \leq 1
\label{ksnrrng}
\end{equation}
such that generically one circle is expanding and the other two are
contracting as one approached the singularity $t\rightarrow 0$.

The distinguishing feature of this class of solutions is that the
spatial slice is flat. In the more generic case, some small elements
of spatial curvature are turned on, rendering the approximation of a
flat spatial slice inconsistent since, as the volume decreases, the
curvature will increase. The details of what happens next were
worked out by BKL for the case of a general homogeneous
3-dimensional space with curvature.

Since the 3 dimensional spatial slice is homogeneous, the metric can
be written as
\begin{align}
ds^2=&  -dt^2 + dl^2 \cr
 dl^2=& \sum_{i=1}^{3} a_i^2(t) (e^i_\mu dx^\mu)\otimes(e^i_\nu dx^\nu)
\end{align}
The 1-forms $e^i_\mu$ are the principal axes of the metric. The
curvature of the slice is encoded in the Maurer-Cartan equations for
these forms, or equivalently by
\begin{equation}
\lambda_i=\epsilon^{\mu_1\mu_2\mu_3}e^i_{\mu_1}\partial_{\mu_2}e_{\mu_3}^i/det(e).
\end{equation}
The classification of possible homogeneous spatial slices, or
equivalently the possible $\lambda_i$, is the Bianchi
classification, which we will not review in a systematic way (a
detailed exposition of this subject can be found at \cite{waldtb}).

Defining the new variables $a_i=e^{\alpha_i}$, Einstein's equations
take on the form
\begin{equation}
\label{bkleqa} 2\alpha_{i,\tau\tau}= {\biggl( \sum_{j\not=i} (-1)^j
\lambda_j a_j^2 \biggr)}^2 -  \lambda_i^2 a_i^4, \end{equation}
\begin{equation}
{1\over 2}\sum_{i=1}^3 \alpha_{i,\tau\tau}=  \sum_{i<j}
\alpha_{i,\tau}\alpha_{j,\tau}
\end{equation}

We will begin with a universe with low curvature, i.e., we will
initially neglect the $\lambda_i$. In this case the RHS of equation
\eqref{bkleqa} is zero and the solution to equation \eqref{bkleqa}
is $\alpha_i = p_i \tau$. Going back to the variable $t$ we obtain
equation (\ref{kasnerb}) i.e, we are back to the Kasner solutions as
expected.

However, as some of the circles shrink, some elements of the
curvature tensor will increase. In equation \eqref{bkleqa}, one of
the $a_i$'s is increasing, say $a_1$ if we choose the conventions in
\eqref{ksnrrng}, and the RHS will be eventually non-negligible.
Keeping only this term in the RHS ($a_{1,2}$ remain consistently
small at this stage) then
\begin{align}
\alpha_{1,\tau\tau}= & -{1\over 2} \lambda_1^2 e^{4\alpha_1},\cr
\alpha_{i,\tau\tau}= & {1\over 2} \lambda_i^2 e^{4\alpha_1},\ \ \
i=2,3
\end{align}
which can be solved explicitly. The net result is the "BKL bounce
rule". The $\alpha_1$ variable, which starts with
$\alpha_{1,\tau}>0$, can be thought of bouncing off an exponential
wall ${1\over2}\lambda_1^2e^{4\alpha_1}$ at large $\alpha_1$. After
being reflected from the wall it will move with $\alpha_{1,\tau}<0$
- i.e. the expanding direction turns into an contracting direction.
The $\alpha_1$ variable moves away from the "curvature wall" and the
effects of this wall become smaller and smaller. Note that for any
value of $\lambda_1$ the behavior of the wall is the same since its
coefficient is $\lambda_1^2$.

While this happens, $\alpha_{2,3}$ which initially satisfy
$\alpha_{2,\tau},\alpha_{3,\tau}<0$ are accelerated such that these
velocities are increased, but again the effects of the "curvature
wall" die off as $\alpha_1$ moves away from the wall. The universe
now enters a new "Kasner free flight" epoch (also known as "velocity
dominated epoch"). The new Kasner exponents can be computed as
described above to be
\begin{multline}
    (p_1,p_2,p_3)\rightarrow \\ \biggl( {|p_1|\over 1-2|p_1|},
    -{2|p_1|-p_2\over 1-2|p_1|},  {p_3-2|p_1|\over 1-2|p_1|} \biggr)
\end{multline}
Now one of the other circles is shrinking.

Another very instructive way to obtain the BKL dynamic is via a
Hamiltonian formulation of GR \cite{misnerbkl}. Going to the
coordinates $(\Omega,\beta_+,\beta_-)$
\begin{align}
\alpha_1=&2(-\Omega+\beta_++\sqrt{3}\beta_-)\cr
\alpha_2=&2(-\Omega+\beta_-+\sqrt{3}\beta_-)\cr
\alpha_3=&2(-\Omega-2\beta_+)
\end{align}
such that $\Omega$ measures the volume of the spatial slice, the
equations of motion are given by the Hamiltonian
\begin{equation}
2{\cal H}=-P_\Omega^2+P_+^2+P_-^2+e^{-4\Omega}(V-1)
\end{equation}
together with a constraint
\begin{equation}
{\cal H}=0.
\end{equation}

$V(\beta)$ is determined by which elements of the spatial curvature
are turned on. The Kasner case (no curvature) corresponds to $V=1$,
and the trajectory is a straight line in $(\Omega,\beta_+,\beta_-)$.
In the case of Bianchi IX space $de^i=\epsilon_{ijk} e^j \wedge
e^k$, which is one the cases in which all the $\lambda_i\not=0$, we
have
\begin{multline}
    V(\beta)=\frac13e^{-8\beta_+} -\frac43
    e^{-2\beta_+}\cosh(2\sqrt{3}\beta_-) + \\\
     +1 +
     \frac23e^{4\beta_+}\biggl(\cosh(4\sqrt{3}\beta_-)-1\biggr).
\end{multline}
\begin{figure}[ht]
\begin{center}
\includegraphics[width=7cm]{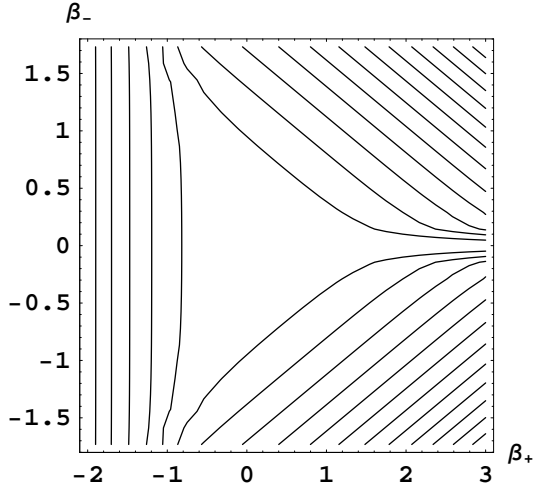}
\caption{\label{BKL1} $V(\beta)$ for Bianchi IX spatial slice}
\end{center}
\end{figure}
The equipotential lines of $V$ are drawn in figure \ref{BKL1}. They
should be thought of as receding to infinity as
$\Omega\rightarrow\infty$. As the space shrinks, $\Omega$ increases
and the potential goes to zero at each fixed $\beta$. The motion
there is approximated by a free Kasner flight. However, if
$(\beta_+,\beta_-)$ are taken to be large enough, then we are
sensitive to the potential. As $(\Omega,\beta_+,\beta_-)$ move in a
straight line, they will eventually catch up with one of the walls
(which are also receding as $\Omega$ increases), bounce off that
wall, and move in a straight line in a different direction - this is
the BKL bounce.

\begin{figure}[ht]
\begin{center}
\includegraphics[width=6cm]{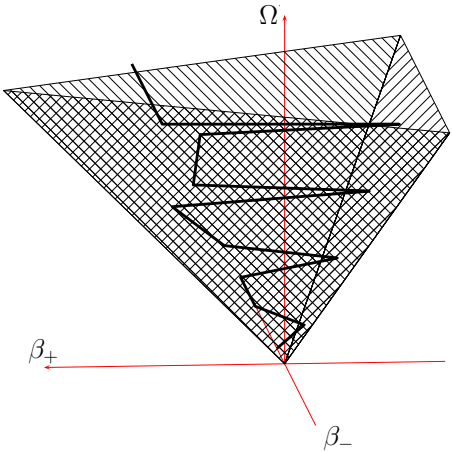}
\caption{\label{BKL2} The BKL trajectory of a spatial slice in
mini-superspace}
\end{center}
\end{figure}

In the regime of large $(\beta_+,\beta_-)$, which is relevant for
small spaces or large $\Omega$, we can truncate to the leading
exponentials in $V(\beta)$. This provides us with 3 sets of walls
with positive coefficients
\begin{equation}
    V\sim \frac13e^{-4\Omega-8\beta_+}+\frac23e^{-4\Omega+4\beta_+}\cosh(2\sqrt{3}\beta_-)
\end{equation}
The walls from which the "Kasner particle" bounces asymptote to the
walls $-4\Omega+8\beta=0$, $-4\Omega+4\beta_+\pm4\sqrt{3}\beta_-=0$.
These form a pyramid in the $\Omega-\beta_+-\beta_-$ space (which is
$\mathbb{R}^{1,2}$), given in figure \ref{BKL2}. The solution is
then described by a particle moving in this wedge, such that
increasing $\Omega$ corresponds to the space shrinking. The
trajectory is a piecewise linear function which bounces off the
walls - this has also been termed "cosmological billiards" since
this system is chaotic. Note, however, that since the volume is
exponential in $-\Omega$ time, the volume of space and its
curvatures reach the Planck scale after only a few bounces. The
analysis then breaks down since it relies on the Einstein Hilbert
action.

For different extensions of GR the structure would be slightly
different - if there are more fields then there would be additional
degrees of freedom beyond $(\Omega,\beta_+,\beta_0)$, there will
other walls etc. In some extensions there are not enough walls to
bound the trajectory to a pyramid and it asymptotes to a straight
line after a finite number of bounces. In other extensions there are
enough walls and the system is chaotic. All stringy cases are of the
latter type. For a review see \cite{Damour:2002et}.

 The solution discussed so far was exact by virtue of
the spatial slice being homogeneous. In the case of a generic
initial condition, i.e., an arbitrary initial spatial slice one can
approximate the behavior around each point by a homogeneous space.
This suggests that each point moves about a pyramid of the type
described above. Of course, the coefficients of the walls will
change from point to point but we see that this has little effect on
the dynamics. However, since the curvature at a point takes into
account only the 1st and 2nd derivative of the metric around a
point, one can then ask what are the effects of higher gradients. It
can be argued that these contribute to subleading walls and hence
also do not change the qualitative dynamics (again the reader is
referred to \cite{Damour:2002et}). This picture is supported by
numerical simulations (for example
\cite{bergerlv,Garfinkle:2003bb}).

Recently, there has been considerable renewed interest in the BKL
dynamics. For a review see \cite{Damour:2002et} and subsequent work
of T. Damour, H. Nicolai, M. Henneaux, A. Kleinschmidt, F. Englert
and others, leading to a very bold conjecture that we will mention
below. The wedge that we identified above in $\mathbb{R}^{1,2}$
turns out to be the Weyl chamber of the hyperbolic algebra $AE_3$
with
\begin{equation}
A_{ij}=    \begin{pmatrix}2 & -1 & 0 \\ -1 & 2 & -2 \\ 0 & -2 &
2\end{pmatrix}
\end{equation}
The submatrix made out of the first two columns and rows is the
Cartan Matrix of $SL(3,R)$, which arises here in the following way.
If we neglect all spatial derivatives then we obtain a reduced GR
action:
\begin{equation}
\int dt {\cal N}^{-1} \biggl( {1\over 4} tr\bigl( (g^{-1}{\dot g})^2
\bigr) - \bigl(tr (g^{-1}{\dot g}) \bigr)^2\biggr)
\end{equation}
where $g$ is the 3-metric on the spatial slice which is $SL(3)$ and
$N$ is the lapse function. The appearance of $AE_3$ is more
mysterious, but it turns out that for all dimensions and extensions
of general relativity, in which the BKL system is chaotic, such an
algebra can be identified such that the BKL motion is given by a
Weyl chamber. For M-theory the algebra is $E_{11}$. This has led
these authors to conjecture that M-theory can be formulated via a
sigma model on this algebra.

\section{Perturbative approaches to spacelike singularities}

\subsection{Quantization of Misner space}

The geometry of Misner space is described in figure \ref{misner}.
Region P (F) describes a big crunch (big bang) and regions L and R
(called "whiskers") are regions with CTC's  - these regions are
Rindler spaces with compact Rindler time. Convenient coordinates for
the different regions are given in \eqref{coora} and \eqref{coorb}.

This model is, in fact, taken to be a "local model" for the
emergence of CTC's in Hawking's chronology protection conjecture
paper \cite{Hawking:1991nk}. Taking region P with metric
$ds^2=-dt^2+t^2dx^2$, one can carry out a change of coordinates
$\tau=t^2,\ v=ln(t)+x$ to bring the metric to the form
$ds^2=-dvd\tau+\tau dv^2$ which covers region P for $\tau>0$ and
region R for $\tau<0$. The symmetries of the space then determine
that a conformally invariant field on this space, times $R^2$, will
have an energy momentum tensor
\begin{equation}
    \langle T\rangle \propto \mathrm{diag}\bigl(t^{-4},3t^{-4},-t^{-4},-t^{-4}\bigr).
\end{equation}
This is one of the motivations for the conjecture that the
divergence of the energy-momentum tensor on the boundary of the
region where CTC's appear will prevent them from forming.

In this section we will discuss the quantization of String theory in
this background as an orbifold, which turns out to be subtle. We
will begin with the simplest sector, which is the untwisted sector,
which already exhibits some unusual features.

\smallskip

 {\bf The untwisted sector}

The untwisted sector states are obtained by symmetrizing flat space
wave functions with respect to the Z action of the orbifold
\cite{Nekrasov:2002kf}, i.e.,
\begin{multline}
\label{untwist}
    f_{j,m^2,s}(x^+,x^-,x)=\int_{-\infty}^\infty dv
    \cr \cdot \exp\Bigl(ip^+X^-e^{2\pi \beta v}+ip^-X^+e^{-2\pi \beta v}
    +\cr + ivj + vs\Bigr)
\end{multline}
where $m$ is the mass of the particle, $s$ is its spin and $j$ is
the boost angular momentum (which is discrete). We will focus on the
case $s=0$ (for a discussion of the precise definition of this wave
function in the case $s\not=0$ see \cite{Berkooz:2004re}).

Note that since the orbifold action does not reverse the orientation
of time, the Misner orbifold maintains the division between positive
and negative energies inherited from flat space. Hence there is no
particle creation in this background. This is similar to the null
brane model \cite{Liu:2002kb,Liu:2002ft}.

Correlation functions for scalars in the untwisted sector states
were calculated in \cite{Berkooz:2002je}. 2 and 3-pt functions are
well behaved but the 4-pt functions diverges as in
\cite{Liu:2002kb,Liu:2002ft} for the null brane. Since each vertex
operator in the untwisted sector is an integral over the flat space
vertex operator, we simply integrate the 4-pt function with 4
integrals of the type \eqref{untwist}. Out of these 4 $v_i$
variables, two are fixed by momentum conservation in terms of the
other two, and one factors out since it is an overall boost of the
entire system, which gives the volume of the angle of the orbifold.
One is left with a single non-trivial integration. If $\Psi_{1,2}$
are the incoming wave functions and $\Psi_{3,4}^*$ the outgoing ones
then the stringy S-matrix (i.e., the integrated 4-pt function) is
\begin{multline}
    <\Psi_1\Psi_2\Psi_3^*\Psi_4^*>=\cr \int_0^\infty
    G(s(v_4))G(t(v_4))G(u(v_4)) \cdot \cr \cdot
    v_2^{il_2+1}v_3^{-il_3+1}v_4^{-il_4-1}\Bigm/\bigm|m_2m_3(v_2^2-v_3^2)\bigm|
\end{multline}
where $v_{2,3}$ are functions of $v_4$ (we will not need the details
of these functions), s,t,u are the Mandelstam variables of the
$2\rightarrow 2$ process that we are discussing, and
$G(x)=\Gamma(-1-x/4)/\Gamma(2+x/4)$.

In the limit $v_4\rightarrow\infty$, $v_2\propto v_4$ and $v_3$ is
constant. We are interested in this limit in order to see whether
the integral diverges. In this limit, one also obtains the relations
$t\rightarrow -({\vec p}_1-{\vec p}_3)^2$ and $s=m_1m_2 v_4$. The
integral then becomes (up to a numerical coefficient)
\begin{multline}
    \propto (m_1m_4)^{1-\frac12({\vec p}_1-{\vec p}_3)^2}
    \left(\frac{m_4}{m_2}\right)^{il_2}
    \left(\frac{m_3}{m_1}\right)^{-il_3}
    \\ \cdot \frac{\Gamma\left(-1+\frac12({\vec p}_1-{\vec  p}_2)^2\right)}{\Gamma(2+t/4)}
    \\ \cdot \int^\infty dv_4\, v_4^{t/2+i(l_2-l_4)}
\end{multline}
which is divergent for $|t|<2$. Qualitatively, it is a UV divergence
that occurs only when distances in the transverse direction are
large enough, and hence has been given the name "IR enhanced UV
divergences" \cite{Liu:2002kb,Liu:2002ft}.

So string theory does not fully remove the divergences. The
divergence is still milder than in GR though. In the limit
$\alpha'\rightarrow 0$, the dependence on $t$ in the integrand
vanishes (since it is really $v_4$ to the power $\alpha' t+...$),
and the integral diverges for all values of $t$. In the limit
$\alpha'\rightarrow 0$ it is also easy to identify that the origin
of the divergence is due to a pole in the t-channel exchange of a
graviton
\begin{equation}
    \int d^8x \int dx^+ dx^- \partial_-\Psi_1\partial_-\Psi_3^*\,\frac{1}{
    \partial^2}\partial_+\Psi_2\partial_+\Psi_4^*
\end{equation}

The effect of a UV divergence associated with larger transverse
separation can be expected on the following grounds
\cite{Horowitz:2002mw}. Consider two particles on the orbifold and
evaluate their interaction by going to the covering space
$\mathbb{R}^{1,1}$, keeping a single image of one of them and
considering its interactions with all the images of the other, which
are related by the boost $\exp(n\beta J_{+-})$ for all integer $n$.
The relative boost between particle 1 and the images of particle 2
therefore increases (indefinitely). Since their center of mass
energy is increasing and the impact parameter does not, then when
the latter is below the Schwarzschild radius of a black hole with
the same mass, the process will be dominated by black hole
formation. The size of black hole in the transverse also grows
larger and larger with $n$, hence we expect a correlation between
the UV divergences in the theory and large distances in the
transverse direction.

\smallskip

{\bf Twisted sector}

We will begin by discussing a seemingly unrelated system - that of a
charged open string in an electric field \cite{Bachas:1992bh} (which
is T-dual to a pair of branes at relative velocity
\cite{Bachas:1995kx}). This system can be realized by taking a pair
of D-branes and turning on an electric field on one of them, or by
turning on an opposite electric field on the two branes. Open
strings in an electric field are given by the boundary conditions
\begin{gather}
    \partial_\sigma X^{\pm}=\mp \pi e \partial_\tau X^\pm\ \ \
    \sigma=0\cr
    \partial_\sigma X^\pm = \pm \pi e \partial_\tau X^\pm\ \ \ \sigma=\pi
\end{gather}
The mode expansion is therefore
\begin{multline}
    X^\pm = x_0^\pm + i \sum_{n=-\infty}^\infty (-1)^n
    \frac{\alpha_n^\pm}{n\pm i\nu} e^{-i(n\pm\nu)}\\
    \cos\left[(n\pm
    i\nu)\sigma\mp i\cdot \mathrm{arcth}(\pi e)\right]
\end{multline}
and the commutation relations are
\begin{gather}
    [a_m^+,a_n^-]=-(m+i\nu)\delta_{m+n},
    \cr [x_0^i,x_0^j]=-i/2e
\end{gather}

Next we need to choose a realization of this algebra, i.e, determine
the Hilbert space and the ground state. The choice for the excited
states $n\not=0$ is straightforward, since the energy of the state
has a real part by which we can divide the operators into creation
and annihilation. The choice for $a_0$ is slightly more complicated.
However, in the limit $e\rightarrow 0$, the $a_0^\pm$ are simply the
momenta of the particle. Hence one realizes these operators as
\cite{Pioline:2003bs} simply as the momentum in the presence of an
electric field
\begin{gather}
\label{opreal}
    a_0^\pm=p^\pm=i\partial_\mp\pm{\nu\over2}x^\pm,\cr
    x_0^\pm\propto i\partial_\mp\mp{\nu\over 2}x^\pm
\end{gather}
and the coordinates $x_0^\pm$ are the coordinates of the center of
the hyperbola on which a charged particle moves in an electric
field.

The Hilbert space is then simply $L^2(\mathbb{R}^{1,1})$. With this
realization the correct normal ordering prescription for the
worldsheet energy-momentum tensor is
\begin{multline}
\label{tnsra}
    L_0=-\frac12(a_0^+a_0^-+a_0^-a_0^+) + \frac{1}{12}(\nu^2-1) + \cr + excited\ states
\end{multline}
which is the just the charged particle Klein-Gordon equation.

Going back to closed strings on Misner space \cite{Pioline:2003bs},
the twisted boundary conditions are
\begin{equation}
    X^{\pm}(\sigma+2\pi,\tau)=e^{\pm
    w\beta}X^{\pm}(\sigma,\tau),\ w\in Z
\end{equation}
give rise to mode expansion
\begin{gather}
    X^\pm_R(\tau-\sigma)=\frac{i}{2}\sum_{n=-\infty}^\infty
    \frac{\alpha^\pm_n}{n\pm i\nu}e^{-i(n\pm i\nu)(\tau-\sigma)} \cr
    X^\pm_L(\tau+\sigma)=\frac{i}{2}\sum_{n=-\infty}^\infty
    \frac{\tilde\alpha^\pm_n}{n\mp i\nu}e^{-i(n\mp i\nu)(\tau+\sigma)}
    \cr
    [\alpha_m^+,\alpha_{-m}^-]=-(m+i\nu),\cr
    [{\tilde\alpha}_m^+,{\tilde\alpha}_{-m}^-]=-(m-i\nu)
\end{gather}
where $\nu=w\cdot \beta$. Following the example of the open string
in electric field we will choose the Hilbert space to be
$L^2(\mathbb{R}^{1,1})$ and the realization of the operators is the
same as in \eqref{opreal} with $a^\pm_0\rightarrow \alpha^\pm_0$ and
$x^0\rightarrow {\tilde\alpha}_0$ (and the $\propto$ sign replaced
by an equality). Both the left and right energy-momentum tensor is
similar to \eqref{tnsra} with $L_0-{\tilde L}_0=2\nu j$ where $j$ is
the quantum number under boosts which is quantized to be an integer.
From now on we will focus on the pseudo zero-modes and neglect the
oscillators (they will play an important role in a computation below
though).

Focusing on the zero modes, the oscillators and mass-shell
conditions translate into equations on the quasi-zero modes of the
form
\begin{equation} M^2=2\alpha_0^+\alpha_0^-,\ \ {\tilde
M^2}=2{\tilde\alpha}^+{\tilde\alpha^-}_0
\end{equation}

For simplicity we will focus on the case $M^2-{\tilde M}^2=0$
(recall that the difference between them is the boost momentum). By
shifting in $\tau$ and boosting (shifting in $\sigma$) we can reach
a situation where the modulus of $\alpha_0^\pm,{\tilde\alpha}^\pm_0$
are all equal. In this case, semiclassically, these equation have
two distinct branches of solutions

\noindent 1) "short strings":
$sign(\alpha^+_0)=sign({\tilde\alpha}^+_0)$ where the classical
string takes the shape
\begin{equation}
    X^\pm(\tau,\sigma)=\epsilon \frac{M}{\nu\sqrt{2}}
    \sinh(\tau)e^{\pm\nu\tau}.
\end{equation}
This is a string that is moving from region P into region F (F into
P) for $\epsilon=1$ ($\epsilon=-1$).

\noindent 2) "long strings" $sign(\alpha^+_0)\not=
sign({\tilde\alpha}^+_0)$, in which case the profile is
\begin{equation}
    X^\pm(\tau,\sigma)=\pm \epsilon \frac{M}{\nu\sqrt{2}}
    \cosh(\tau)e^{\pm\nu\tau}.
\end{equation}
For $\epsilon=+1$ (-1) this is a string which wraps the time
direction in whisker R (L), and in the radial direction comes in
from infinity to some finite distance from the origin and then goes
out to infinity again.

Next we would like to ask whether these strings, and primarily the
"long strings", do anything interesting. One can either condense
them or ask whether they are created in pairs - if one starts from a
configuration in which the twist quantum number is conserved then
only the latter is allowed. Next, one can try and evaluate how such
strings backreact on the geometry, an issue which is not well
understood (see however \cite{Berkooz:2005ym}). Here we will briefly
go over one of the arguments that these strings are pair produced,
after carefully considering the 2nd quantization (in space-time) of
the model \cite{Berkooz:2004re}.

We have seen that there are two types of strings - "short" and
"long". The short strings have a unique vacuum - strings propagating
forward in time are creation operators and those backwards in time
are annihilation operators. For the long strings the ground state is
ambiguous. This can be understood as follows: the closed strings
quasi-zero-modes can be viewed as two copies of the momentum of a
charged particle in an electric fields, or we can take the left
moving quasi-zero-modes to be momenta and the right movers to be the
center the hyperbola, or we can switch the role of the left and
right movers. Each choice like this has a different natural vacuum.
In all these vacua there is either twisted sector particle pair
creation, just as particles in an electric field, or a large
degeneracy of vacua which one needs to sum over.

{\bf 3-pt function}:

One way of possibly evaluating the effects of condensation of
twisted modes on the geometry is to try and use conformal
perturbation theory. Although we may not expect that this will make
the model completely non-singular, it might give us some
information. The first step is to evaluate a 3-pt function of 2
twisted sector states and one untwisted. This will also give us
information on how the regular graviton sees the twisted sector
states, ie., what is their profile of $\langle T_{\mu\nu}\rangle$.
This was carried out in \cite{Berkooz:2004yy}.

The simplest way to carry out the computation of a 3-pt function
with two twisted sector vertex operators and a single untwisted
vertex operator is to insert the twisted operators at
$\tau\pm\infty$ on a cylinder, and the untwisted vertex operator
somewhere in between, which means computing a 1-pt function of an
untwisted operator in the twisted sector. We will focus on a
universal stringy part of the result, which is due to the excited
string modes. The latter turns out to be reminiscent of
non-commutative geometry.

The untwisted vertex operator is defined in the untwisted sector as
\begin{equation}
    V_T=\lim_{w\rightarrow z}
    e^{ik^+X^-(w)}e^{ik^-X^+(z)}e^{k^+k^-\log(w-z)}
\end{equation}
When we insert this operator in the twisted sector, the simplest way
to carry out the computation is to re-normal-order it with respect
to the twisted sector vacuum:
\begin{multline}
    V_T=e^{i\left(k^+X^-_{\prec 0}+k^-X^+_{\prec 0}\right)}
    e^{i\left(k^+X^-_{\succ 0}+k^-X^+_{\succ 0}\right)}
    \\
    e^{i(k^+X^-_0+k^-X^+_0)}
    e^{k^+k^-\left([X_{>0}^-,X_{<0}^+]-[X_{\succ 0}^-,X^+_{\prec
    0}]\right)}
\end{multline}
where $X_\prec$ ($X_\succ$) is the creation (annihilation) part of
$X$ with respect to the twisted vacuum, and $X_>$ and $X_<$ the
similar parts with respect to the untwisted vacuum.

The last exponential is simply
\begin{equation}\label{krnl}
    e^{-k^+k^-\left(\psi(1+i\nu)+\psi(1-i\nu)-2\psi(1)\right)}.
\end{equation}
At large values of $\nu\sim\beta w$, where $w$ is the twisted sector
number, the coefficient of $k^+k^-$ is $-2\cdot\ln(\nu)$, i.e.,
arbitrarily large. For any wave function that we take for a twisted
sector state $\Psi(x^+,x^-)$, ordinary untwisted string states will
see them smeared (by an amount which can be very large) by the
kernel \eqref{krnl} acting on $|\Psi|^2$. This non-locality is
similar to the one seen in non-commutative geometry - wave functions
which have higher $k^+$ momentum will be more delocalized in the
$k^-$ direction.

\section{Tachyon condensation}

we have focused before on twisted sector modes in the general Misner
orbifold. Let us know specialize to the case that the rate in which
the orbifold decreases is small, i.e, $\beta<<1$, or conversely
${\dot R}<<1 $ in region P where we have a circle of radius R
shrinking to zero \cite{McGreevy:2005ci}.

In this case, when the radius of shrinking circle reaches about the
String scale, and if the fermions have anti-periodic boundary
conditions around the orbifold (a Scherk-Schwarz compactification
\cite{Scherk:1978ta}), the lowest winding state becomes tachyonic.
Since we have assumed that ${\dot R}=\beta<<1$ at $R\sim l_s$, the
singularity is still a long time ahead in the future, $l_s/\beta$,
and we can deal with the effects of the condensation of this tachyon
without worrying about the singularity \cite{McGreevy:2005ci} (more
complicated collapsing geometries are discussed in
\cite{Silverstein:2005qf} and application to the "final state of the
black hole" \cite{Horowitz:2003he} are discussed in
\cite{Horowitz:2006mr}).

The condensation of a tachyon in a time direction is not understood
as much as the condensation of tachyon on a time-like singularity,
i.e., a tachyon localized along a spatial direction. Nevertheless,
we will use the latter to get some intuition. Furthermore, tachyon
condensation in closed strings is not as well understood as in open
strings (for a review see for example \cite{Sen:2004nf}), but the
general features will suffice for our purposes.

In \cite{Adams:2001sv} it was shown that non-supersymmetric orbifold
singularities relax to flat space (or to supersymmetric orbifold
singularities) by condensing twisted sector closed string tachyons.
Geometrically, the orbifold looks like a cone, whose tip is resolved
by String theory. In the case that the singularity relaxes to flat
space, the condensation of the tachyon removes the tip of the cone
and smoothes it out (when the tip of the cone has been smoothed to
below string scale curvatures one can use GR to describe its
relaxation to flat space). We see that the condensation of a
localized tachyon excises the part of space where the tachyon
profile is localized. This can be understood generally from
worldsheet considerations - the condensation of the a tachyon
decreases the central charge on the worldsheet (remember that we are
discussing the static spatial part of the target space, which is
unitary) and therefore removes degrees of freedom. In particular it
can remove the degrees of freedom that correspond to string states
moving in that region of space - hence parts of space are removed.
Another class of localized tachyons is discussed in
\cite{Adams:2005rb}.

A similar process can be exhibited in the context of the AdS/CFT
duality. The Hawking-Page phase transition is understood to be a
confinement/de-confinement phase transition in the dual field theory
\cite{Witten:1998zw}. In \cite{Barbon:2001di} it was shown that this
phase transition can be understood as a condensation of a localized
Atick-Witten tachyon, which is easy to identify if thermal AdS is
superheated (tachyons play additional important roles in AdS - for
example see \cite{Horowitz:2005vp,Ross:2005ms} and references
therein). Another clean model in the context of closed strings is
the FZZ duality between $SL(2)/U(1)$ (the Euclidean two dimensional
black hole) and sine-Liouville theory \cite{fzz} (more information
on this duality can be found at \cite{Kazakov:2000pm} and subsequent
work. The supersymmetric case is discussed in
\cite{Giveon:1999px,Giveon:1999tq}). The $SL(2)/U(1)$ background is
\begin{gather}
    ds^2=k(dr^2+\tanh^2r d\theta^2),\ \
    \theta\sim\theta+2\pi\cr
    \Phi=\Phi_0-2\ln\bigl(\cosh(r)\bigr)
\end{gather}
i.e, it is a space which starts as an $S^1\times$(linear dilaton) at
infinity and caps off by pinching the $S^1$ at some finite value of
the dilaton. The sine-Liouville background is given by the 2D
lagrangian
\begin{multline}
    L= \frac1{4\pi}\biggl( (\partial x)^2 + (\partial \phi)^2 + Q{\hat
    R}\phi +\\
    +\lambda e^{-Q/b}\cos \left[R(X_L-X_R)\right] \biggr).
\end{multline}
In this case there is no geometric end at the strong coupling
region, but rather there is a winding mode, similar to the tachyon
we have discussed above, whose profile increases in the strong
coupling region. Since the two models are the same quantum
mechanically, we see that the main effect of this winding mode
condensate is to excise the region of space where it condenses.

In the context of a time dependent background and a state becoming
tachyonic at a given moment in time and onwards, we expect that the
condensation of this tachyon will cut the time direction
\cite{McGreevy:2005ci}. In the case that we are discussing, since
$\beta<<1$, this happens long before one reaches the singularity and
hence one needs not worry about it or about the large blue shift
effects associated with it. The hope is that this new kind of cap
will have finite stringy amplitudes.

There are two approaches that one can take. In both of them one
starts with an ordinary spacelike Liouville theory
\begin{equation}
    S=\int \frac{d^2\sigma}{4\pi} \biggl( (\partial\phi)^2+\mu e^{2b\phi}
    \biggr)
\end{equation}
with a background charge $Q=b+1/b$. One can then either use the
Euclidean solution, whose correlation functions are finite and well
understood, as a kind of Hartle-Hawking state \cite{Hartle:1983ai},
which is the approach taken in
\cite{McGreevy:2005ci,Silverstein:2006tm,Silverstein:2005qf,Nakayama:2006gt},
Or one can try and solve it in Minkowski space
\cite{Strominger:2003fn,Fredenhagen:2003ut,Horowitz:2006mr}.

Both approaches have some intriguing features, but are not without
open problems. The idea that the condensing tachyon is a uniquely
stringy Hartle-Hawking cap is very appealing. But since the problem
is inherently Minkowskian it relies on the assumption that the
analytic continuation is well defined, whereas it is clear that it
is very subtle. Also, using a Hartle-Hawking state requires that one
will be able to glue the Euclidean cap to a Minkowskian spacetime
such that we will inherit a Minkowskian initial condition. It is not
clear how to do so for Liouville theory which does not have a
$\phi\rightarrow -\phi$ symmetric slice.

Studying the Minkowskian setup directly also produces some unusual
features. If one attempts to define it via the analytic continuation
of the Euclidean Liouville with $\phi=iX^0$ and $\beta$ purely
imaginary then one can show that it does not satisfy all axioms of
CFT \cite{Strominger:2003fn} (certain 3-pt functions do not reduce
to 2-pt functions upon insertion of the identity operator as one of
the 3 operators). A direct quantization of the model in Minkowski
space has been carried only in mini-superspace thus far
\cite{Fredenhagen:2003ut}. The Minkowskian model
\begin{align}
{\cal L}=(\partial_\tau X^0)^2 + \mu e^{\beta X^0},&& \mu>0
\end{align}
describes a particle moving in an inverted exponential potential.
Such a particle reaches $X^0=\infty$ at finite time. Hence one needs
to supplement boundary conditions at infinity (or more radically one
can attempt to glue it to another space altogether). Already in
minisuperspace these boundary conditions are not unique, but rather
parameterized by a single real number, and it remains to be seen
what happens in the case of a full field theoretic analysis.

For further reading on recent progress in the understanding of
time-like and null tachyons the reader is refereed to
\cite{Aharony:2006ra,Hellerman:2006hf,Hellerman:2006ff,Hellerman:2006nx}.

\section{Non-perturbative methods}

\subsection{Using the AdS/CFT}

As we discussed in section 2, inside the Schwarzschild black hole
there exists a spacelike singularity. In this section we will
describe some of the efforts to understand such singularities in the
context of black holes in AdS, using the AdS/CFT correspondence.

\subsubsection{Simulation of black holes, microstates, and the state at the BH singularity}

The statement of the AdS/CFT correspondence \cite{Maldacena:1997re}
is that a field theory, which is an object that we understand rather
well quantum mechanically, is completely equivalent (or more
precisely dual via a strong/weak coupling duality) to a
gravitational background. A thermal state in the field theory is
dual to a non-extremal black hole on the gravitational side. The
most straightforward way therefore to probe a black hole is simply
to run a simulation of the thermal state in the field theory.

In practice, a full simulation is not doable since we need to take
the large N limit to obtain a weakly curved gravitational
background. However a mean field simulation of the field theory for
D0-branes (i.e, quantum mechanics) was carried out in
\cite{Iizuka:2001cw,Kabat:2001ve,Kabat:2000zv,Kabat:1999hp} with
reasonable agreement to the black hole behavior in the gravity dual
to D0-branes \cite{Itzhaki:1998dd}.

One can then ask whether one sees any remanent of the singularity.
The answer is that it is difficult to see any remnant of the region
behind the horizon at all. The black hole is described by a thermal
state, and probe computations, with which spacetime is defined in
this framework, breakdown at the horizon - a D0 brane probe which is
brought close to the cluster of thermalized D0 branes will simply
thermalize once it reaches the vicinity of the horizon. This
thermalization is signaled by the appearance of a tachyon in the
off-diagonal mode between the probe D0 brane and one of the D0
branes in the thermal state.

In this picture the whole region behind the horizon is replaced by
the thermal state. This picture is reminiscent of the micro-state or
fuzzball picture of the black hole advocated by Mathur (for a review
see \cite{Mathur:2005zp}). In this picture the black hole is
replaced by an ensemble of classical GR solution without horizons.
These solutions are very similar in the region outside the would-be
horizon but are very different inside it where they develop a
multi-throat cap. In this description the singularity also
disappears (and in fact spacetime is regular, singularity free and
horizon free throughout the solution). The black hole is considered
to be, in such a picture, an effective description of the ensemble
of such states. The caveat is that actually the different
configurations differ only on very small scales, which means that
under any small perturbation they will start mixing extensively,
which might indicate that the ensemble picture is always the correct
description. Still it could be that for the study of the singularity
one needs to go back to the micro-state language - to our knowledge
this has not been done.

A related development is the suggestion by Horowitz and Maldacena
\cite{Horowitz:2003he} that for the purposes of computing in the low
energy effective action, one needs to place a specific state at the
spacelike singularity. This might make the evolution from past null
infinity to future null infinity unitary in the region outside the
black hole. To make contact with the discussion before, since the
singularity is in the future of all observers behind the horizon,
one can take the state at the singularity and evolve it backwards in
time to a state at the horizon, i.e., we end up with a description
in which we have cut space at the horizon and replaced it by some
new complicated state. Actually, if the thermal state description in
the D0 quantum mechanics could be clarified, or how to sum over the
GR microstates, this could be a way to calculate the state at the
black hole singularity.

\subsubsection{The eternal black hole in AdS spaces}

The Penrose diagram of the eternal black hole in $AdS_d$ ($d > 3$)
is given in figure \ref{adsbh} \cite{Fidkowski:2003nf} (the case d=3
is special and is described in \cite{Kraus:2002iv}). This
configuration is dual to the thermofield description of finite
temperature field theory \cite{Maldacena:2001kr}. The state in the
field theory which is dual to the black hole is the thermal density
matrix
\begin{equation}
    \rho=\sum_n e^{-\beta E_n} |n><n|
\end{equation}
where the sum over $n$ is a sum over energy eigenstates which span
the physical Hilbert space ${\cal H}$. Note, however, that the
eternal Minkowski black hole has two boundaries which are $S^3\times
R$. This geometry is encoded by going to the thermofield
description, which is a different way of writing the density matrix
above. In this description one uses a pure state in a doubled
Hilbert space ${\cal H}\otimes {\cal H}$ (with hamiltonian
$H=H_1-H_2$)
\begin{equation}
    |\Psi\rangle =\sum_n e^{-\beta E_n/2}|n\rangle_1*|n\rangle_2\in {\cal H}*{\cal H}
\end{equation}
up to a normalization. A correlation function of operators $O(t_i)$,
which act on the first copy of ${\cal H}$, in this state satisfies
\begin{equation}
\langle\Psi| O(t_1)...O(t_n)|\Psi\rangle=tr\biggl(\rho\cdot
O(t_1)..O(t_n)\biggr)
\end{equation}
(at different times and positions, when we are suppressing the
latter).

In the thermofield description, however, we now have the option to
act with operators on either Hilbert spaces. Let us take two
operators $A(t_1)$ and $B(t_2)$ and denote on which Hilbert space
they act by using an additional lower index 1 or 2. An operator
$O_1(t)$ ($O_2(t)$) acts on the first (second) copy of ${\cal H}$.
In this case, the following relation holds
\begin{multline}
    \langle\Psi|A_1(t_1)B_2(t_2)|\Psi\rangle=\\
    =\langle\Psi|
    A_1(t_1)B_1(t_1-i\beta/2) |\Psi\rangle.
\end{multline}
Hence there is an easy interpretation for computing correlators with
insertions on both boundaries of the eternal BH.

The question is whether we can compute some correlators on the
boundary that will probe the singularity. We will then try and
evaluate these correlators in the dual field theory, and see how the
singularity changes as $g_s$ is taken to be non-zero (finite N), or
when string worlsheet corrections are taken into account (smaller 't
Hooft coupling).

Whether we can find such correlators is not a-priory clear. On the
one hand the Minkowskian black hole is the analytic continuation of
the Euclidean black hole. In the latter, the background terminates
at the horizon, well before reaching the singularity. This
background in itself contains most of the physical information about
the thermal field theory, leading one to expect that we will not be
able to observe anything behind the horizon. On the other hand, if
one inserts two operators on the two boundaries then the propagator
probes the region behind the horizon and one can attempt to probe
the singularity this way \cite{Fidkowski:2003nf,Kraus:2002iv}.
\begin{figure}[ht]
\begin{center}
\includegraphics[width=6cm]{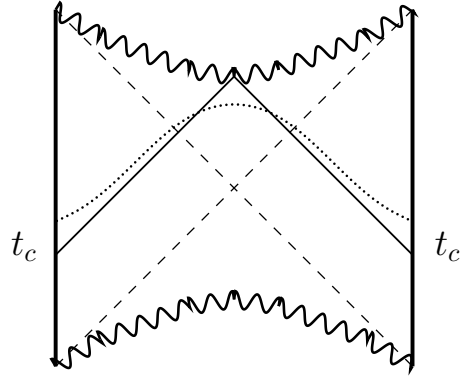}
\caption{\label{adsbh2} The dotted line is a geodesic going from one
boundary to the other. The solid line is the critical null geodesic
which touches the singularity.}
\end{center}
\end{figure}

Focusing on very massive particles, the propagator is dominated by a
single space-like geodesic which goes through the horizon. For
$AdS_d$ $d>3$ one can choose points on the two boundaries such that
the geodesic between them passes as close as we want to the
singularity, and its proper length in the interior is as small as we
want (i.e., it approaches a null geodesic). In fact there are two
critical points, each one at time $t_c$ on each of the boundaries,
such that the geodesic between these points is null and touches the
singularity (figure 8). The existence of such a geodesic implies
that $\langle\Phi(t_c)\Phi(t_c-i\beta/2)\rangle$ diverges. So it
seems that there is a signature for the singularity
\cite{Fidkowski:2003nf}.

This raises a puzzle. The fact that the two point function diverges
is actually in contradiction to an inequality in finite temperature
field theory
\begin{equation}
|\langle\Phi_1(t)\Phi_2(-t)\rangle | \ <\
\langle\Phi_1(0)\Phi_2(0)\rangle \ <\ \infty
\end{equation}
This inequality indicates that in the Minkowskian black hole, viewed
as an analytic continuation of the Euclidean black hole, one is not
supposed to take this geodesic into account. This is a situation
which is known to happen when evaluating integrals (in this case, a
path integral) via the steepest descent method - the largest saddle
point need not dominate if it is not on a good steepest descent
contour. Hence, it seems that again we do not see the singularity.

However, this geodesic, i.e saddle point in the Green's function, is
still there. We can analytically continue time in this Green's
function even further then the $\pi/2$ rotation that takes us from
Euclidean to Minkowski space such that in this new non-physical
regime, the null geodesic will be the dominant contribution. Hence,
there will be a pole in the correlation function in some
non-physical sheet. In principle one can try and see if such a pole
exists and study its behavior as N and the 't Hooft coupling are
varied. In \cite{Fidkowski:2004fc} preliminary results were
presented that the pole does not exist at large N, weak 't Hooft
coupling (using an approximation of the correlator which is then
analytically continues to the non-physical sheet). This may indicate
that the singularity is smoothed out already by $\alpha'$ effects.
However, this assumes that the procedure of extracting the pole is
reliable, which is not a settled issue yet. A detailed description
of the analytic structure of the finite temperature field theory,
some more observables that one can construct and a discussion of
what it might take to measure them with enough precision is given in
\cite{Festuccia:2006sa,Festuccia:2005pi}.

\subsection{M(atrix) models}

 A BFSS \cite{Banks:1996vh} type Matrix model for a linear
dilaton background in a null direction, $g_s=e^{-Qx^+}$, was
discussed in \cite{Craps:2005wd} and \cite{Craps:2006xq}.
Transforming to the Einstein  frame this gives us a big-bang
configuration. The M(atrix) model turns out to be 1+1 SYM with a
varying Yang-Mills coupling in the field theory. Since the string
coupling becomes strong (weak) at the early (late) times, the YM
coupling is weak (strong) there \cite{Dijkgraaf:1997vv}.
Equivalently by a scale transformation, one can describe the model
as a 1+1 SYM with fixed coupling on a worldsheet which looks like
the future cone of Misner space. A M(atrix) model for the null brane
was discussed in \cite{Robbins:2005ua} and \cite{Martinec:2006ak},
and it also turns out to be a time-dependent lagrangian on the
worldvolume of the matrix theory. A D-Instanton probe of the
null-brane singularity was done in \cite{Berkooz:2005ym}.

These models share several features, so we will discuss them
together. We will begin with some caveats and then focus on the
physics that can be extracted from the models. The first caveat is
that both models rely on a perturbative analysis of the closed
string background, and do not allow for changing the background in a
large way. This is worrisome since we believe that these backgrounds
are unstable as they have large divergences. Another issue, which
was addressed in the papers, is that the theory on the D0 or D1
branes breaks supersymmetry. In such cases very often a
non-supersymmetric brane configuration generates tadpoles for closed
string modes and one needs to condense these modes, go to the stable
point, and only then decouple the theory on the open branes.
Furthermore, without SUSY there is no guarantee that the
perturbative computation will give the same results as the large N,
very low energy, (1/N), computation required for the BFSS
conjecture.

Out approach will be the following. First of all, even without using
a BFSS like conjecture, these models provide information on the
behavior of stringy probes near the singularity. Second, if relying
on a BFSS like conjecture, we will view the results of the M(atrix)
analysis as indicating some general features of the behavior near
the singularity, rather than a full qualitative computational tool.

The result that is obtained in both models is that near the
singularity, the full $N*N$ degrees of freedom of the matrices need
to be used. Usually when the full $N*N$ degrees of freedom are
excited (and assumed in a thermal phase) one obtains the M(atrix)
description of a black hole \cite{Banks:1997hz,Banks:1997tn}. For
the null brane this has a clear interpretation in the spirit of the
Polchinski-Horowitz effect - as the D0 branes (or gravitons in a
BFSS interpretation) are squeezed together near the singularity,
they form black holes and the full set of degrees of freedom of the
black holes have to be used. In the case of the linear dilaton
background, the interpretation is less clear but seems to indicate a
new non-geometrical phase at the big bang singularity (perhaps
reminiscent of the "gas of black holes" or "maximally stiff equation
of state" hypothesis of
\cite{Banks:2006hy,Banks:2004cw,Banks:2002fe,Banks:2001px}).

\section{ Acknowledgement }

It was a pleasure talking with many people on these subject - O.
Aharony, B. Craps, B. Durin, S. Elitzur, F. Englert, A. Giveon, Z.
Komargodsky, D.Kutasov, H. Liu, B. Pioline, G. Rajesh, E.
Rabinovici, V. Schomerus, S. Sethi, S. Shenker, E. Silverstein and
J. Simon. This work is supported by the Israel Science Foundation,
by the Braun-Roger-Siegl foundation, by EU-HPRN-CT-2000-00122, by
GIF, by Minerva, by the Einstein Center and by the Blumenstein
foundation.

\end{document}